\documentstyle[preprint,epsfig,eqsecnum,aps]{revtex}
\def\btt#1{{\tt$\backslash$#1}}
\begin{document}
\draft
%\preprint{HEP/123-qed}
\title{Description of collective magnetic states within a spherical single 
particle basis}
\author{A. A. Raduta$^{a)}$, A. Escuderos$^{b)}$ and E. Moya de Guerra$^{b)}$\\}
\address{
$^{a)}$Department of Theoretical Physics and Mathematic, Bucharest University,
POB MG6, Romania\\
Institute of Physics and Nuclear Engineering, Bucharest POB MG6, Romania\\}
\address{$^{b)}$ Instituto de Estructura de la Materia, CSIC, Serrano 119-123, 28006 Madrid, Spain}
\date{\today}
\maketitle
\begin{abstract}
A many body Hamiltonian comprising pairing, quadrupole-quadrupole and spin-spin interaction is treated within a projected spherical basis with the aim of describing the detailed structure of the magnetic states of orbital and spin-flip nature in $^{144, 148-154}$Sm. The mean field for the single particle motion takes care of the volume conservation of a phenomenological core which results in having a complex
dependence on deformation for single particle energies. 
Several experimental features like the total orbital and total spin
strengths, the shape of the orbital strength distribution, the hump structure of the spin strength distribution, the quadratic dependence of the orbital strength on nuclear deformation, the  saturation effect of the orbital strength with respect to the Casten's factor P are very well described within a quasiparticle random phase approximation formalism. Comparison of present results with those
obtained in some previous
treatments are also given.

\end{abstract}
\pacs{PACS number(s): 21.10.Re,~~ 21.60.Ev,~~27.10.+j,~~27.70.+q}
%{\tt$\backslash$\string pacs\{\}} should always be input,
%even if empty.}

\narrowtext

\section{Introduction}
\label{sec:level1}

The collective magnetic dipole mode (M1) was one of the central subjects in the last two decades.
The interest has grown with time due to the wealth of experimental data provided by several groups performing either $(e,e^{\prime})$ experiments at backward angles with high resolution or $(\gamma,\gamma^{\prime})$ measurements using brehmstralung radiation.
The theoretical work in this field preceded the experiment. Indeed 
Lo Iudice and Palumbo \cite{Iu} predicted the existence of this state from an
anti-phase angular oscillation of proton and neutron axially symmetric and rigid systems. The model is known under the name of two rotor model (TRM), and the predicted mode as scissors mode. Later on Iachello predicted the existence of such a mode at lower energy within the Interacting Boson Approximation (IBA)
\cite{Iache}. The experimental discovery 
of this state in $^{156}$Gd, by Achim Richter and his collaborators
\cite{Boh}, in an $(e,e^{\prime})$ experiment, was a stimulus for both experimental
\cite{Ber,Wes,Zie,Rang,Pit,Mar,ARich}
 and theoretical
\cite{Roh,Rad0,Rad2,Rad3,Gin,Fas2,Fas1,Fes,Za1,Hil,Iu1,Za2,Sug,Iu2,Ga,Sar,Ham,Mag,Hey,Cos1,Cos2,Zaw,Zaw1,IBA2,Rad9,Rad1}
 intensive activities. Thus, many descriptions have been proposed and some authors reactivated some old  own papers devoted in the past to closely related topics.
Both phenomenological\cite{Roh,Rad0,Rad2,Gin,Fas2,IBA2} and microscopic 
\cite{Rad3,Fas1,Fes,Za1,Hil,Iu1,Za2,Sug,Iu2,Ga,Sar,Ham,Mag,Hey,Cos1,Cos2,Zaw,Zaw1,IBA2,Rad9,Rad1}
descriptions have been proposed.
The phenomenological formalisms provide an intuitive basis for the description
of the new mode, although many details such as  the fragmentation and the magnitude of the total strength cannot be accounted for. That is the reason why several improvements have been proposed along the time. The rigid motion of the proton-neutron system was replaced by a share motion\cite{Hil} or the motion of two liquid drops\cite{Roh,Fas1}.  The Interacting Boson Approximation was extended by distinguishing between the proton and neutron bosons \cite{IBA2} and pointed out that the F spin symmetry plays an important role for the structure of the magnetic state. Also the Coherent State Model\cite{Rad6} was generalized to  GCSM\cite{Rad0}, which provides a  new boson description of M1 states.
The microscopic formalisms use the QRPA approach within a deformed single particle basis and predict indeed that the M1 strength is distributed among many QRPA states. The nice feature brought by the microscopic theories is
that the low lying states with relatively large M1 transition probability
are excited mostly by the orbital part of the transition operator. Such type of states are conventionally called  scissors like states since their structure confirm the hypothesis of the TRM model that the mode is due to the convection current created by the protons motion.

Early microscopic description of the state as a doorway state orthogonal to the spurious rotational modes predicted fragmentation
\cite{Rad3,Rad9,Dip,Moy,Pov,Rad8}
The fragmentation effect was confirmed experimentally and
 as a consequence a new  question was opened, namely,  whether there are 
orbital states in the region beyond 4 MeV where the spin-flip excitations dominate. A long standing debate was about the mixture in the predicted mode, of spurious components of isoscalar nature\cite{Zhe}. Several projection procedures were proposed\cite{Fas1,Hil,Ham,Dip,Moy,Noj,Moy1} to purify the theoretical state from spurious components.

Many review articles discussing the results on this topic appeared in journals or proceedings of conferences and summer schools.
Since we focus our attention on a specific description, we might unavoidably omit
some important contributions to the field from our list of references
and therefore, we advise the reader to consult the most recent reviews from 
Refs.\cite{Zaw3,Rich,Iu3}.

The present paper is devoted to the study of the orbital and spin excitations in the even-even isotopes of Sm, using  a projected single particle basis \cite{Rad5}. Some ingredients like the projected states  and the model Hamiltonian are similar to those used in Ref.\cite{Rad1}. The new point of the present investigation consists
of the definition of the mean field. Here the mean field of the single particle motion is defined by averaging a particle core Hamiltonian on a coherent state
describing a phenomenological deformed core. The coupling term comprises a monopole-monopole and a quadrupole-quadrupole interaction. The monopole term is calculated by exploiting the restriction for the volume conservation. In this way the single particle energies, approximated by the average of the mean field Hamiltonian on the projected states, have a complex dependence on the deformation parameter. We recall that in Ref. \cite{Rad1} the single particle energies depend linearly on deformation.
Therefore our aim is to investigate whether by considering  more realistic 
single particle energies one obtains an improvement of the quantitative description given in Ref. \cite{Rad1}. As we shall see at the end of this paper many details shown by the experimental data are nicely described by our approach which in fact confirm our expectation that in order to describe the non-global features of the system one should define properly the mean field of the single particle motion.

The results are described  according to the following plan.
In Section 2, we present briefly the projected single particle basis. Also, here we describe the mean field and derive analytical expressions for the single particle energies. The model Hamiltonian and the QRPA states are introduced in Section 3. The reduced probability for exciting these states from the ground state is  also given. Applications to the even-even isotopes of Sm are described in Section 4. The main results are summarized in Section 5

\section{Projected single particle basis}
\label{sec:level2}
Technical details about the construction of the angular momentum projected single particle basis 
appropriate for the description of the single particle motion in a 
deformed mean field generated by the particle core interaction, are given in 
Ref. \cite{Rad1}. Aiming at a self-consistent presentation, we give here the main ingredients of the method used in the above quoted reference.

The single particle mean field is determined by a particle-core Hamiltonian: 
\begin{equation}
\tilde{H}=H_{sm}+H_{core}-
M\omega_0^2r^2\sum_{\lambda=0,2}\sum_{-\lambda\le\mu\le\lambda}
\alpha_{\lambda\mu}^*Y_{\lambda\mu}.
\end{equation}
Here $H_{sm}$ denotes the spherical shell model Hamiltonian while $H_{core}$ is
a harmonic quadrupole boson ($b^+_\mu$) Hamiltonian associated to a phenomenological core. The interaction of the two subsystems is accounted for by 
the third term of the above equation, written in terms of the shape coordinates $\alpha_{00}, \alpha_{2\mu}$.
The quadrupole shape coordinates are related to the quadrupole boson
 operators by the canonical transformation:
\begin{equation}
\alpha_{2\mu}=\frac{1}{k\sqrt{2}}(b^{\dagger}_{2\mu}+(-)^{\mu}b_{2,-\mu}),
\end{equation}
where $k$ is an arbitrary C number. The monopole shape coordinate is to be 
determined from the volume conservation condition. In the quantized form, the result is:
\begin{equation}
\alpha_{00}=\frac{1}{2k^2\sqrt{\pi}}\left[5+\sum_{\mu}(2b^{\dagger}_{\mu}b_\mu
+(b^{\dagger}_{\mu}b^{\dagger}_{-\mu}+b_{-\mu}b_{\mu})(-)^{\mu})\right].
\end{equation}
Averaging $\tilde{H}$ on the eigenstates of $H_{sm}$, hereafter denoted by
$|nljm\rangle$, one obtains a deformed boson Hamiltonian whose ground state is described by a coherent state
\begin{equation}
{\Psi}_g=exp[d(b_{20}^+-b_{20})]|0\rangle_b,
\end{equation}
with $|0\rangle_b$ standing for the vacuum state of the boson operators and $d$ a real parameter which simulates the nuclear deformation.
On the other hand, the average of $\tilde{H}$ on ${\Psi}_g$ is  similar to the Nilsson Hamiltonian \cite{Nil1}.
Due to these properties, it is expected that the best trial functions 
to be used to generate, through projection, a spherical basis are:
  
\begin{equation}
{\Psi}^{pc}_{nlj}=|nljm\rangle{\Psi}_g.
\end{equation}
The upper index appearing in the l.h. side of the above equation suggests that the product function is associated to the particle core system
The projected states are obtained in the usual manner by acting on these deformed states with the projection operator
\begin{equation}
P_{MK}^I=\frac{2I+1}{8\pi^2}\int{D_{MK}^I}^*(\Omega)\hat{R}(\Omega)
d\Omega.
\end{equation}
A certain subset of projected states is orthogonal:

\begin{equation}
\Phi_{nlj}^{IM}(d)={\cal N}_{nlj}^IP_{MI}^I[|nljI\rangle\Psi_g].
\end{equation}
The main properties of these projected spherical states are:
a) They are orthogonal with respect to I and M quantum numbers.
b) Although the projected states are associated to the particle-core system,
they can be used as a single particle basis. Indeed when a matrix element of a particle like operator is calculated, the integration on the core collective coordinates is performed first,  which results in obtaining a final 
factorized expression: one factor carries the dependence on deformation and
one is a spherical shell model matrix element.
c) The connection between the nuclear deformation and the parameter $d$ 
entering the definition of the coherent state (2.4) is readily obtained by requiring that the strength of the particle-core quadrupole-quadrupole interaction 
be identical to the Nilsson deformed term of the mean field:
\begin{equation}
\frac{d}{k}=\sqrt{\frac{2\pi}{45}}(\Omega^2_{\perp}-\Omega^2_z).
\end{equation}
Here $\Omega_{\perp}$ and $\Omega_z$ denote the frequencies of Nilsson's mean field related to the deformation $\delta=\sqrt{45/16\pi}\beta$ by:
\begin{equation}
\Omega_{\perp}=(\frac{2+\delta}{2-\delta})^{1/3},~
\Omega_{z}=(\frac{2+\delta}{2-\delta})^{-2/3}.~
\end{equation}
The average of the particle-core Hamiltonian $H'=\tilde{H}-H_{core}$
on the projected spherical states defined by Eq.(2.7) has the expression
\begin{eqnarray}
\epsilon_{nlj}^I&=&\langle\Phi_{nlj}^{IM}(d)|H'|\Phi_{nlj}^{IM}(d)\rangle=\epsilon_{nlj}-\hbar\omega_0(N+\frac{3}{2})C_{I0I}^{j2j}C_{1/201/2}^{j2j}
\frac{(\Omega^2_{\perp}-\Omega^2_z)}{3}\nonumber\\
&+&\hbar\omega_0(N+\frac{3}{2})\left[1+\frac{5}{2d^2}+\frac
{\sum_{J}(C^{jIJ}_{I-I0})^2I^{(1)}_J}{\sum_{J}(C^{jIJ}_{I-I0})^2I^{(0)}_J}\right]
\frac{(\Omega^2_{\perp}-\Omega^2_z)}{90}.
\end{eqnarray}
Here we used the standard notation for the Clebsch Gordan coefficients
$C^{j_1 j_1 j}_{m_1 m_2 m}$.  $I_J^{(k)}$ stands for the following integral
\begin{equation}
I^{(k)}_J=\int^{1}_{0}P_J(x)[P_2(x)]^kexp[d^2P_2(x)]dx, ~k=0,1
\end{equation} 
It is worth mentioning that the norms for the core's projected states as well as the matrix elements of any boson operator on these projected states can be
fully determined once the overlap integrals defined in (2.11), are known
\cite{Rad6}.
Since the core contribution does not depend on the quantum numbers of the 
single particle energy level, it  produces a shift for all energies and therefore is omitted in 
Eq.(2.10). However when the ground state energy variation against deformation is studied, this term should be considered as well.

The first term from (2.10) is, of course, the single particle energy for the spherical shell model state $|nljm\rangle$. 
According to our remark concerning the use of the projected spherical states for describing the single particle motion, the average values
$\epsilon_{nlj}^I$ may be viewed as approximate expressions for the single
particle energies in deformed Nilsson orbits. We may account for the deviations from the exact eigenvalues by considering, later on, the exact matrix elements of the two body interaction when a specific treatment of the many body system is applied.
It is worth mentioning that the dependence of the new single particle energies on deformation is similar to that shown by the Nilsson model. This is clearly
seen in Fig. 1 and Fig. 2 described in section 4.
We remark that the energies shown in the above mentioned plot depend on deformation in a different manner than those obtained by one of the present authors in Ref. \cite{Rad1}. Indeed therein, they depend linearly on deformation while here non-linear effects are present. The difference between the two sets of energies is caused by the fact that here the volume conservation condition was used for the monopole shape coordinate while in Ref.\cite{Rad1} this term was ignored
and moreover there is no account of the volume conservation condition.
In this respect, one expects that this difference in the single particle energies will induce some modifications of the results of Ref.\cite{Rad1} concerning the fragmentation process of the M1 strength.
Although the energy levels are similar to  those of the Nilsson model, the 
quantum numbers in the two schemes 
are different. Indeed here we generate from each j a multiplet of $(2j+1)$ 
states as I, which plays the role of the Nilsson quantum number $\Omega$, runs from 1/2 to j.
On the other hand, for a given I there are $2I+1$ degenerate sub-states while the Nilsson states are only double degenerate. As explained in Ref.\cite{Rad1}, the redundancy 
problem can be solved by changing the normalization of the model functions:
\begin{equation}
\langle\Phi_{\alpha I M}|\Phi_{\alpha I M}\rangle=1 \Longrightarrow \sum_{M}\langle\Phi_{\alpha IM}|\Phi_{\alpha IM}\rangle=2.
\end{equation}
Due to this normalization, the states $\Phi_{\alpha IM}$ used to calculate the 
matrix elements of a given operator should be multiplied with the weighting factor $\sqrt{2/(2I+1)}$.

Concluding, the projected single particle basis is defined by Eq.(2.7). The
projected states are thought of as eigenstates of an effective Hamiltonian, with the corresponding  energies given by Eq.(2.10). 

\section{The model Hamiltonian and the M1 states} 
\label{sec:level3} 
In the states introduced in the previous Section, we consider a system of Z protons and N neutrons interacting among themselves in the following manner. The nucleons of similar isotopic charge interact through pairing interaction,  dipole-dipole and quadrupole quadrupole interaction. Protons and neutrons interact
only through dipole and quadrupole two body interaction. The many body Hamiltonian describing such a system is written in second quantization as:

\begin{eqnarray}
H&=&\sum_{\alpha IM;\tau}\epsilon_{\alpha I}(\tau)c^+_{\alpha IM}(\tau)c_{\alpha IM}(\tau)-\sum_{\tau}\frac{G_{\tau}}{4}P_{\tau}P^+_{\tau}
\nonumber\\
&-&
\frac{1}{2}\sum_{\mu\tau\tau'}X_{\tau\tau'}^{(1)}\sigma_{\mu}(\tau)
\sigma_{\mu}^+(\tau')-\frac{1}{2}\sum_{\mu\tau\tau'}X^{(2)}_{\tau\tau'}
Q_{2\mu}(\tau)Q^+_{2\mu}(\tau').
\end{eqnarray}
Note that although the model Hamiltonian is formally identical to that used in Ref.\cite{Rad1}, the mean field parts of the two schemes are different from each other.
Therefore even if the procedures used are similar, the final results 
are different.
Before applying the QRPA treatment, it is convenient to rewrite the 
QQ interaction by re-coupling the initial and final nucleons in a different way and keep in the final result only the dipole-dipole term. By doing this, the new expression is no longer separable.
The many body Hamiltonian is written first in quasiparticle representation, by using the Bogoliubov-Valatin transformation, defined by the coefficients $U_{\alpha I}$ and $V_{\alpha I}$.
At this stage one can define the QRPA dipole phonon operator which creates, by acting on the vacuum state $|0\rangle $, the QRPA states: 

\begin{equation}
|1^+\mu ,\lambda \rangle=\sum_{a<b;\tau}[X_{ab}^{\lambda}(\tau)
A_{\mu}^+(ab;\tau)-
Y_{ab}^{\lambda}(\tau)A_{\tilde{\mu}}(ab;\tau)]|0\rangle\equiv\Gamma_{1\mu}^+(\lambda)|0\rangle,
\end{equation}
where $A_{\mu}^+(ab;\tau)$ and $A_{\tilde {\mu}}(ab;\tau)$ denote the dipole operator constructed out of two quasiparticle creation operators $a_{\alpha}^+$ and $a_{\beta}^+$ and two annihilation operators   $a_{\alpha}$ and $a_{\beta}$,
respectively 	
\begin{equation}
A_{\mu}^+(ab;\tau)=\sum_{\alpha \beta}C_{m_{\alpha}m_{\beta}\mu}^{I_a I_b 1}a_{\alpha}^+a_{\beta}^+,
~~~ A_{\tilde {\mu}}(ab;\tau)=\left (A_{-\mu}^+(ab;\tau)\right )^{\dagger}(-)^{1-\mu}.           
\end{equation}
The phonon amplitudes are obtained by solving the QRPA equations with the
normalization condition:

\begin{equation}
\sum_{a<b;\tau}\{|X_{ab}^{\lambda}|^2-|Y_{ab}^{\lambda}|^2\}=1.
\end{equation}
The solutions of the QRPA equations are labeled by the index $\lambda =1,2,...$
defined by the ordering relation for the corresponding energies
($\omega_{\lambda}$):
$\omega_1<\omega_2<\omega_3<...$ 
The QRPA states can be excited from the ground state, the vacuum state of the phonon operator, by the transition operator:
\begin{equation}
{\cal M}(M1,\mu)=\sqrt{\frac{3}{4\pi}}\sum_{i,\tau}[g_l^{eff}(\tau)l_{\mu}(i,\tau)+g_s^{eff}(\tau)s_{\mu}(i,\tau)].
\end{equation}
where the effective gyromagnetic factors are obtained by accounting for the
renormalization induced by the angular momentum carried by the core system:
\begin{equation}
g^{eff}_l=g_l-g_c,~g^{eff}_s=g_s-g_c.
\end{equation}
The reduced transition probability is defined by\footnote{Throughout this paper the Rose convention for the reduced matrix element is used}:
\begin{equation}
B(M1;0^+\rightarrow 1_{\lambda}^+)\equiv B(M1\uparrow)=|\langle 0||{\cal M}(M1)||1^+,\lambda\rangle|^2.
\end{equation}
The involved matrix element has the expression:
\begin{eqnarray}
\langle 0||{\cal M}||1^+,\lambda\rangle&=&\sqrt{\frac{3}{\pi}}\sum_{a<b;\tau}\frac{1}{\hat{I}_b}(g_l^{eff}(\tau)l_{ab}(\tau)+
\frac{1}{2}g_s^{eff}(\tau)\sigma_{ab}(\tau))
\nonumber \\
&\times& (U_a(\tau)V_b(\tau)-U_b(\tau)V_a(\tau))[X_{ab}^{\lambda}(\tau)-Y_{ab}^{\lambda}(\tau)],~~ {\hat I}=\sqrt{2I+1}.
\end{eqnarray}
with obvious notations for the reduced matrix elements for the orbital angular momentum (l) and spin operator ($\frac{1}{2}$).

\section{Application to the even-even Sm isotopes}
\label{sec:level4}
The QRPA equations were solved for $^{144,148-154}$Sm isotopes.
The single particle energies are those given by Eq. (2.10), with the spherical
shell model energies $\epsilon_{nlj}$ determined by the frequency $\omega_0$
and the strengths C and D for the spin orbit and $l^2$ terms, respectively
\begin{equation}
\omega_0=41A^{-1/3}MeV,~~C=-2\omega_0 \chi,~ D=-\omega_0\chi \mu,
\end{equation}
where $\chi=0.0637,~\mu=0.6$ for protons and $\mu=0.42$ for neutrons
\cite{Nil1,Nil2}.
The single particle energies can be considered either as functions of the nuclear 
deformation
$\delta$ by means of the frequencies $\Omega_{\perp}$, $\Omega_z$, or as  functions of the deformation parameter $d$ due to  Eq. (2.8). In the later situation,
an additional parameter is introduced,$k$, which remained undetermined when
the shape coordinates were transformed in boson variables. In order to  
compare the present results with those reported in Ref.\cite{Rad1} it is useful to define the strength of the particle core qQ interaction so that the corresponding matrix elements have similar expressions in the two descriptions. It results
\begin{equation}
X_c=\frac{\hbar \omega_0}{8k}\sqrt{\frac{5}{2\pi}}.
\end{equation}
For illustration, proton and neutron single particle energies are presented 
as functions of the deformation parameter $d$ in the figures 1 and 2, 
respectively, for $^{154}$Sm. The particle-core
coupling constant $X_c$ was taken equal to that given in Ref.\cite{Rad1}.
In principle the deformation parameter should be determined by requiring that 
the system of Z protons and N neutrons filling the energy levels in Figs. 1 
and 2 respectively, in accordance to the Pauli exclusion principle, has a minimum energy with respect to changing the deformation. By this procedure, one obtains the equilibrium value of $d$ specified by the dashed vertical line. However in the 
generalized coherent state model (GCSM)\cite{Rad0}, which is also based on coherent states description, the deformation parameter was 
fixed by a fit of the experimental E2 strength  to the lowest $2^+$ states. It turns out that this procedure works well for deformed nuclei but not 
unambiguously for near spherical nuclei. The value obtained for the case of 
$^{154}$Sm, by using this method, is mentioned by vertical full lines 
in Figs. 1, 2. The deviation of this value from the equilibrium deformation might be corrected in two ways; either by modifying the strength $X_c$ (or, equivalently, the parameter $k$) so that the two values are the same, or by considering the normalization of the single particle energies due to the motion of the core. Indeed,
we recall that the term of the model Hamiltonian which corresponds to the core motion was set to zero when the average on the coherent state describing the phenomenological core was performed. The average of the core Hamiltonian is a function of $d$ which influences, to a large extent, the equations for energy minimum.
The Fermi energies for protons and neutrons are pointed by arrows in the corresponding graphs.

The strength parameters of the model Hamiltonian which were used in our calculations are collected in table 1. As we have already mentioned, we want to outline the effect on the M1 strength coming from the fact that we use a different set of single particle energies than in Ref.\cite{Rad1}. 
 Such an effect could be best determined if the strength parameters
 used in the two descriptions were close to each other. Indeed the parameters from Table 1 are, with few exceptions, practically the same as in Ref.
\cite{Rad1}.
We modified the pairing strengths so that the gap parameters produced with the new set of single particle energies were the same as in the previous treatment.
Moreover these are the same as those used by Hamamoto and Magnusson in 
Ref\cite{Ham}.
Only for $^{144}$Sm and $^{154}$Sm one notices a deviation of the proton gap of about 100keV. Also the strength of the QQ interaction for $^{150}$Sm and
$^{152}$Sm is smaller in the present case by a factor of about 2.
The strengths of the pairing interaction are consistent with the results obtained by Nilsson \cite{Nil1} while the strengths of dipole and quadrupole interactions are close to the parameters used by de Coster and Heyde \cite{Hey,Cos1,Cos2} and by Hamamoto and Magnusson\cite{Ham}. The results of the BCS treatment are shown in terms of quasiparticle energies for $^{154}$Sm, in Figs. 3 and 4. From these figures, one sees that despite the fact that the  considered nucleus is a deformed one, the quasiparticle energies exhibit a shell like structure manifested by some energy gaps.
The quasiparticle energies were used to solve the QRPA equations. Further, the 
transition operator given by Eq. (3.5) is used to excite the resulting states.
The reduced transition probability for such excitation to take place is calculated by means of relation (3.8) which contains in a factorized form the contribution coming from the BV and QRPA transformation amplitudes as well as from the matrix elements of the orbital and spin operators.
%\begin{table}[ht]
%\caption

\vskip0.2cm

\begin{tabular}{lllllllllll} \hline
    & $d$ & $k$ & $4X_c$ & $G_p$ & $\Delta_p$ & $G_n$ & $\Delta_n$ &
    $-X^{(1)}_{pp}$ & $X^{(1)}_{pn}$ & 
    $b^4 X^{(2)}_{pp}$ \\ 
    & & & & & & & & (MeV) & (MeV) & (MeV) \\ \hline

$^{144}$Sm & 0.400 & 2.330 & 1.500 & 20.70/A & 1.592 & 19.51/A & 0.951 & 
    0.797 & 0.0171 & 1.196 \\
$^{148}$Sm & 0.500 & 2.750 & 2.800 & 19.66/A & 1.328 & 17.35/A & 1.060 & 
    0.600 & 0.0166 & 1.164 \\
$^{150}$Sm & 1.482 & 3.770 & 1.100 & 20.52/A & 1.227 & 18.16/A & 1.331 & 
    0.600 & 0.0164 & 0.533 \\
$^{152}$Sm & 1.850 & 4.980 & 0.688 & 18.27/A & 0.865 & 17.08/A & 1.049 & 
    0.500 & 0.0162 & 0.658 \\
$^{154}$Sm & 2.290 & 5.580 & 0.611 & 18.20/A & 0.777 & 15.80/A & 0.721 & 
    0.850 & 0.0160 & 1.118 \\ \hline
    
\end{tabular}

%\end{table}
\small{Table 1.The parameters entering the many body Hamiltonian (3.1) are listed. Gap energies for proton and neutron systems are also given. The strength of the QQ interaction is multiplied by the fourth power of the oscillator length 
$b$(=$\sqrt{\hbar/(m\omega_0)}$.

\vskip0.2cm
The results for M1 strength are presented in several figures for the energy regions of orbital and spin transitions.
The total strengths characterizing the intervals 2-4 MeV and 4-10 MeV
are listed in Table 2. Comparing the data given there and the corresponding ones given in Ref. \cite{Rad1} one immediately concludes that the single particle energies have a notable effect on both the orbital and the spin excitations in the even-even Sm isotopes.
For the gyromagnetic factors of orbital angular momenta, we took their bare
values, i.e. $g_l(p)=1$ and $g_l(n)=0$. The spin gyromagnetic factors are quenched by a factor 0.85, as in Ref.\cite{Ham}, in order to account  for effects determined by nucleon excited states which are not described in the present paper.
For the gyromagnetic factor induced by  the core angular momentum, we took the value $g_c=0.166$ which corresponds to a ratio of the renormalized neutron to proton orbital gyromagnetic factors equal to 0.2, as suggested by the phenomenological models IBA2 \cite{IBA2}  and GCSM \cite{Rad11} .   Our analysis suggests a small contribution due to the core gyromagnetic factor. For orbital transition strength this confirms the scissor character of the transition, i.e. it is described by an operator proportional to 
$(g_p-g_n)(l^p_\mu-l^n_\mu )$. As far as the spin-flip transitions are 
concerned, the correction due to $g_c$  is small since this is small comparing it to the spin gyromagnetic factor.
The theoretical and experimental spin-spin strength distributions are compared with each other in Fig. 5. It is interesting to remark that all peaks in the two panels showing experimental and theoretical results can be easily put in one to one correspondence. The difference between the two pictures consists of that two of 
our predicted peaks at about 7.2 and 7.8 MeV are higher than the corresponding experimental  peaks 7.5 and 7.9MeV. Also in the low part, the experience shows a broad peak of 0.05 $\mu_N^2$ while in the present calculations one finds a narrow peak which is 0.4 $\mu_N^2$.
\vskip0.2cm
\begin{tabular}{cccccc} 
\hline
   $\hskip1cm$ &  $\hskip1cm$ $^{144}$Sm $\hskip1cm$ &  $\hskip1cm$ $^{148}$Sm
 $\hskip1cm$ 
& $\hskip1cm$ $^{150}$Sm $\hskip1cm $& $\hskip1cm$ 
 $^{152}$Sm$\hskip1cm$ & $^{154}$Sm \\
    \hline
$S_1$ & 0.260 & 0.401 & 0.954 & 2.324 & 2.508 \\
$S_2$ & 5.656 & 5.473 & 7.484 & 7.755 & 7.961 \\ 
\hline    
\end{tabular}
\vskip0.2cm
\small{Table 2. Predictions for the total orbital strength, below 4 MeV, and total 
spin-flip transitions to the states $1^+$ lying in the range of 4-10 MeV
are listed under the labels $S_1$ and $S_2$ respectively. For the case of 
$^{154}$Sm the transition to the state at 4.08 MeV is included in $S_1$(see text). }

\vskip0.2cm
The distribution of the orbital strength predicted for the QRPA states lying below the limit of 4 MeV is compared with the experimental data in Fig. 6.
Several features  revealed by this picture should be mentioned. The number of states shown by 
 experiment is equal to  the corresponding theoretical number. Exception is $^{152}$Sm where the first two small experimental sticks correspond to one relatively big stick. Even in this case, the dominant 4 peaks bunched around 3 MeV are well simulated by the predicted peaks in the range of 2.9-3.6 MeV. Although the main peaks are shifted a little with respect to the corresponding experimental results, the shape of the theoretical and experimental strength distribution are  in good agreement with each other.
Also we notice that going from the deformed nucleus $^{154}$Sm toward the spherical
case of $^{148}$Sm, the orbital strength is less fragmented which is consistent with the fact that deformation causes fragmentation.
The transition probabilities to all states from the interval 2-10 MeV are shown in Fig. 7. While in $^{154}$Sm the strength is almost continuously distributed and shaped in maxima alternating with minima, in the near spherical nuclei the interval containing strength is decreased and moreover the groups of peaks are well separated and less members are included in a group. Several doublet structures are noticed in the lightest two isotopes. 
From our analysis it results that one cannot state that there is a sharp separation between orbital and spin like excitation and that the border is 4 MeV.
On the contrary, one could say that for deformed case there are reasonable chances to find orbital transition beyond 4 MeV while for lighter isotopes the spin transition could be visible below 4 MeV. For example in $^{154}$Sm the transitions to the states of energy equal to 3.69 and 4.03 MeV are of proton spin nature 
while the state at 5.419 MeV is of orbital nature. Indeed for the latest transition the ratio orbital to spin  is 2.11 and its magnitude is 
0.33 ${{\mu }_N}^2$.
In $^{152}$Sm the states with the energies 2.79 and 3.64 MeV have  spin-neutron and spin-proton character. $^{150}$Sm has also spin-flip like states below 4 MeV, at 3.05 MeV, and an orbital state of energy equal to 4.005 MeV.
In our formalism none of the states below 4 MeV calculated for $^{144,148}$Sm is of pure orbital nature. Indeed, for $^{148}$Sm the two states lying below 4 Mev in Fig 7 are characterized by the ratio $R_{o/s}=0.54;0.26$ while for 
 $^{144}$Sm this ratio is equal to 0.324. In all three cases the spin contribution is brought by protons. The highest peak in the heaviest three isotopes is of neutron nature and is lying in the interval 7.5-8MeV. The second highest peak in $^{154}$Sm is dominated by the proton spin component but has also neutron admixture. Its energy is  smaller than that of the neutron peak, by 700keV. In the next two isotopes there is no second peak of height comparable to that of the neutron peak. In $^{148}$Sm there are two proton peaks with energies 4.05 and 7.51 MeV with strength slightly larger than 0.5$ {{\mu}_N}^2 $ and a neutron peak at 5.5 MeV with a strength of about 0.5 ${{\mu}_N}^2$. The lightest isotope exhibits a dominant neutron peak at 6.35 MeV, two proton peaks lying close to 4.5 and 9 MeV and three neutron peaks with energies 5.3, 5.4 and 6.6 MeV. Our calculations suggest that one cannot separate sharply an energy interval where only neutron spin strength can be found and one only with proton transitions. However a certain polarization of the strength distribution may be seen. For example, for the first three nuclei ($^{154,152,150}$Sm) beyond 7.5 MeV transitions are, with very few exceptions of small proton transitions, of neutron type. In the last two ( $^{144,148}$Sm) cases in the interval mentioned above, the situation is opposite. An interesting feature one sees in $^{144}$Sm is that the regions with proton and neutron like transitions come in successive order. Indeed in the interval 4-5 MeV one observes only one proton peak. The next interval 5-7MeV is populated mainly by neutron transitions, the protons appear to be impurities being the smaller components in the last two doublets.   
Finally, beyond 7 MeV there are only proton like transitions.

One of the most interesting properties of the scissors like excitation for Sm isotopes was pointed out in Ref.\cite{Rang}, and refers to the linear dependence of the total strength below 4 MeV, on the deformation squared.
This dependence is nicely obtained in Fig. 8 and this suggests once more
 a very good agreement with the experiment. Such a relationship between the total orbital strength and nuclear deformation is a rigorous prove that
the magnetic states populated in the inelastic (e,e$^{\prime}$) backward scattering are of collective nature. On the other hand, the E2 transition probability from the ground state to the first collective $2^+$ state has also a linear dependence on the nuclear deformation squared. This suggests a proportionality relationship between the two quantities. Indeed in Ref.\cite{Cast1,Cast2} it has been shown that the
proportionality factor is  the average energy defined as
\begin{equation}
{\bar E}_x=\sum {E_iB_i(M1,\uparrow)}/\sum {B_i(M1,\uparrow)}.
\end{equation}
The theoretical and experimental average energies,  defined as above, 
are compared with each other in Fig. 9. The agreement is again very good.
Indeed the deviation of theory from experimental data vary from 100keV
in the case of $^{154}$Sm,  to 210 keV in $^{144}$Sm. 
The collective features of the scissors states might be also evidenced in 
the plot of the total orbital strength below 4 MeV versus the factor 

\begin{equation}
P = \frac{N_pN_n}{N_p+N_n} ,
\end{equation}
introduced by Casten in Ref. \cite{Cast1}. Indeed, 
as shown in Ref. \cite{Cast2} the graph obtained is almost identical to that representing the BE2 values as function of the same quantity.
In the above equation $N_p, N_n$ denote the proton and neutron boson numbers respectively, defined in the IBA model as being equal to half the nucleon number in the open shell.
In our case the predicted and experimental values  were interpolated as
in the reference quoted above by a continuous curve. We note that here the saturation process start with $^{152}$Sm at $P\approx 5$ and the limit strength is 2.9
$\mu_N^2$. 

\section{Conclusions}
\label{sec:level5}
A model Hamiltonian has been treated within the QRPA formalism based on a projected spherical single particle basis. The mean field for the single particle motion is defined by averaging a particle-core Hamiltonian over a coherent state describing a deformed phenomenological core. The monopole part of the coupling term is determined from the volume conservation condition.

An orthogonal basis is defined through projection technique from a product function
of a spherical single particle state and a coherent state for the quadrupole shape coordinate describing the phenomenological core.
Although this basis comprises both the single particle and core coordinates, it may be used for treating the operators acting only on particle coordinates by
integrating on the core coordinates.
A set of single particle energies are defined by averaging the mean field Hamiltonian on each member of the basis. These energies depend on deformation in a similar way as the Nilsson energies. It is argued that these energies approximate quite well the eigenvalues of the mean field Hamiltonian while the projected states are close to the states which are obtained by projecting the good angular momentum  from the eigenstates of the mean field Hamiltonian.

The QRPA formalism provides the dipole magnetic states. Special attention is focused to those states lying in the interval 0-10 MeV. For these states the
B(M1) values were calculated by using the transition operator given by Eq. (3.5).

Applications were made for the even-even isotopes of Sm.
The intervals 2-4 and 4-10 MeV were considered separately since the states in these intervals are excited by different parts of the transition operator.

For the first interval we compared the predictions with the corresponding data for the following observables. The distribution of the theoretical orbital 
strength has similar shape as the experimental results. Moreover the number of the transitions seen experimentally is equal to the number of transitions
of a visible size in the plots shown. 

The very good quality of the agreement between  theory and experiment can be seen from the plot showing the quadratic dependence on the nuclear deformation of the total orbital strength. The same conclusion  can be drawn from the plots
representing the average energy in the interval 0-4 MeV as well as from the
graph of total strength as function of the Casten's factor P.

Concerning the spin transitions  we considered first the case of $^{154}$Sm
where experimental results for strength distributions are available. As shown in Fig. 5, the predicted  distribution agrees quite well with the experimental one.
For all considered isotopes the spin transitions are presented in Fig. 7.
One clearly notices how the fragmentation is influenced by deformation and moreover the appearance of the shell structure when one goes from the deformed to the spherical isotopes. Comments about the isotopic dependence of the spin strength and upon the fact whether there are orbital transitions beyond 4 MeV or spin transitions below 4 MeV are included.

We stress that the formalism used in the present paper is similar to that from Ref.\cite{Rad1} except for the single particle energies. Indeed there they have a linear dependence on deformation while here higher powers of deformation are included by implementing the volume conservation condition for the mean field. This feature seems to be essential in obtaining a better description of both the orbital and spin strengths.

As a final conclusion one may say that the present model, using a projected 
spherical single particle basis and a mean field defining a set of deformed single particle energies which depend non-linearly on deformation, is an efficient
method for a realistic description of the M1 properties caused by convection  
currents and spin distributions in nuclei.

{\bf Acknowledgments.} Part of this work was supported by CNCSIS under contract 
A918/2001, Romania, and by PB98/0676 CICYT, Spain.

\vfill\eject

\begin{figure}[h]
\centerline{\psfig{figure=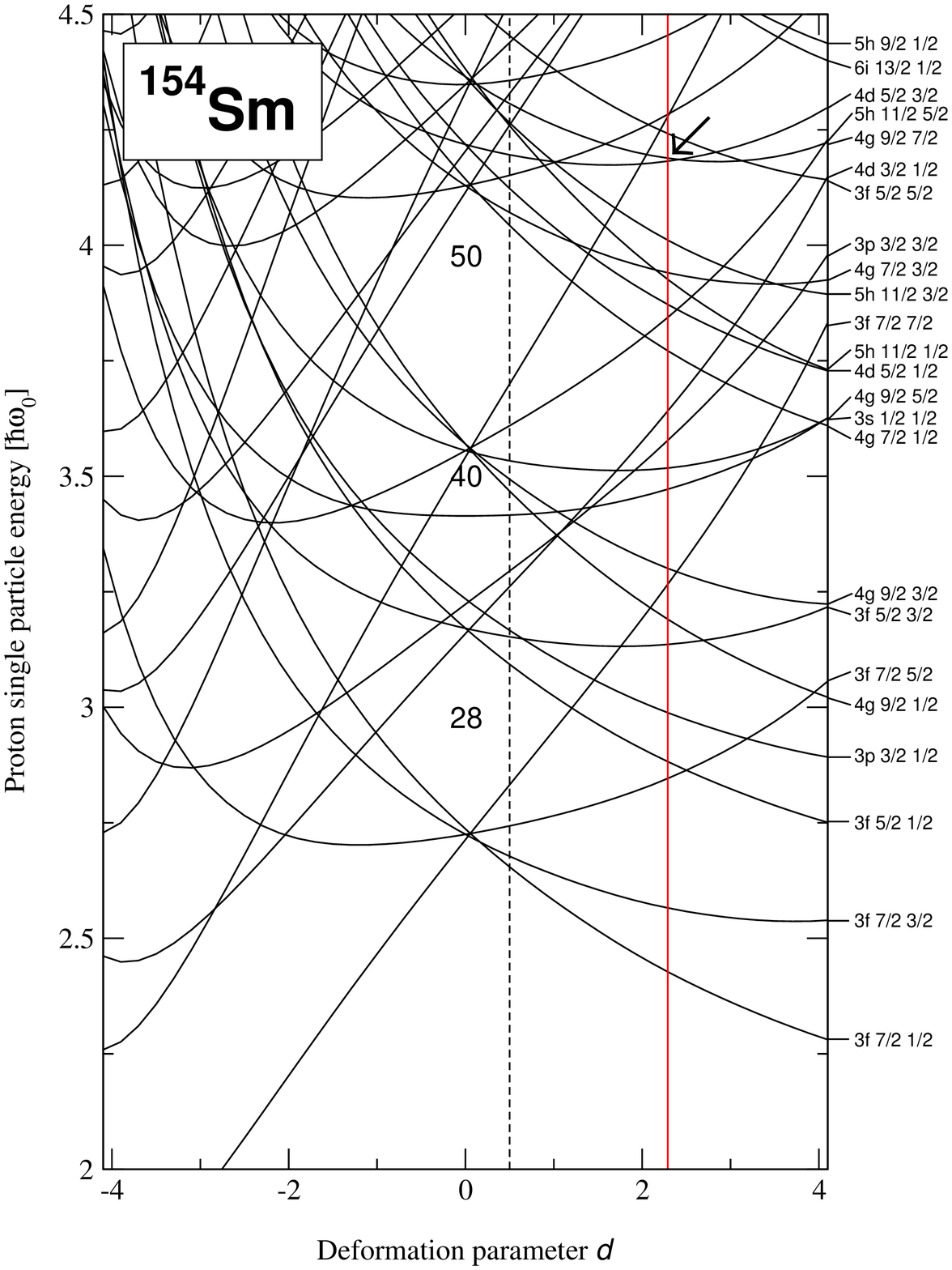,width=8cm,bbllx=5cm,%
bblly=10cm,bburx=18cm,bbury=26cm,angle=0}}
\vskip8cm
\caption{Proton single particle energies given in units of $\hbar\omega_0$.}
\label{Fig. 1}
\end{figure}
\clearpage

\begin{figure}[h]
\centerline{\psfig{figure=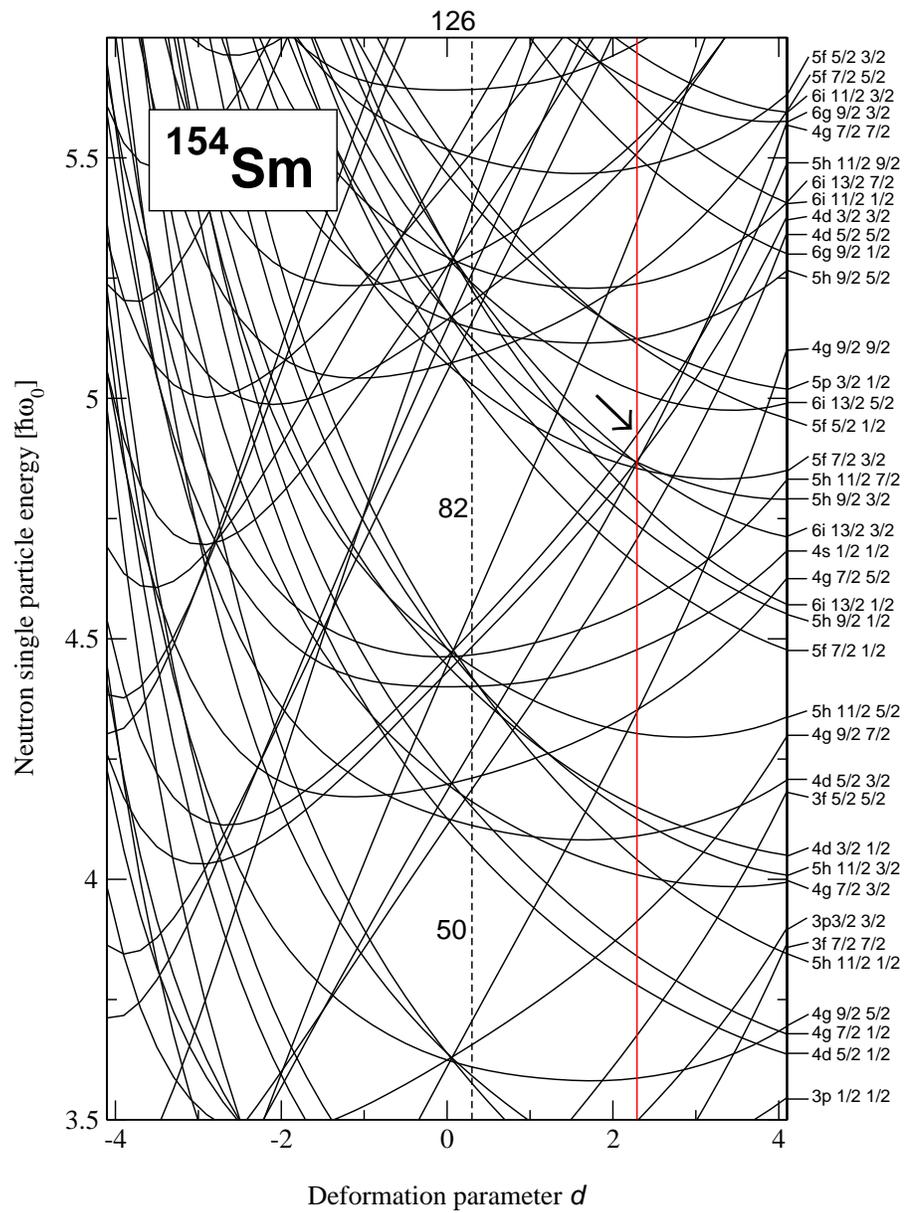,width=8cm,bbllx=5cm,%
bblly=10cm,bburx=18cm,bbury=26cm,angle=0}} 
\vskip8cm
\caption{Neutron single particle energies given in units of $\hbar\omega_0$.}
\label{Fig. 2}
\end{figure}
\clearpage

\begin{figure}[h]
\centerline{\psfig{figure=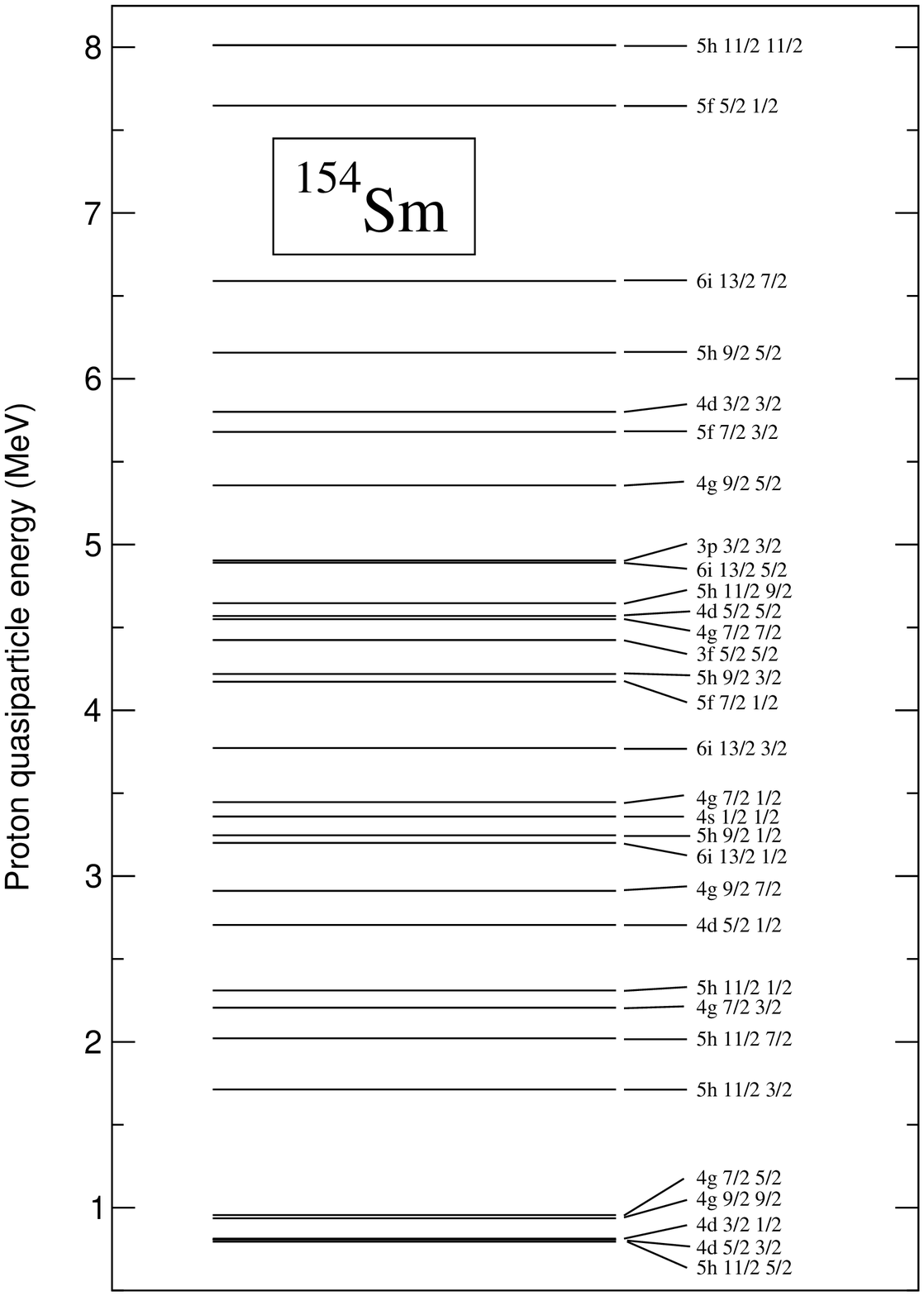,width=10cm,bbllx=5cm,%
bblly=10cm,bburx=18cm,bbury=26cm,angle=0}}
\vskip8cm
\caption{Neutron quasiparticle energies given in units of MeV.}
\label{Fig. 3}
\end{figure}
\clearpage

\begin{figure}[h]
\centerline{\psfig{figure=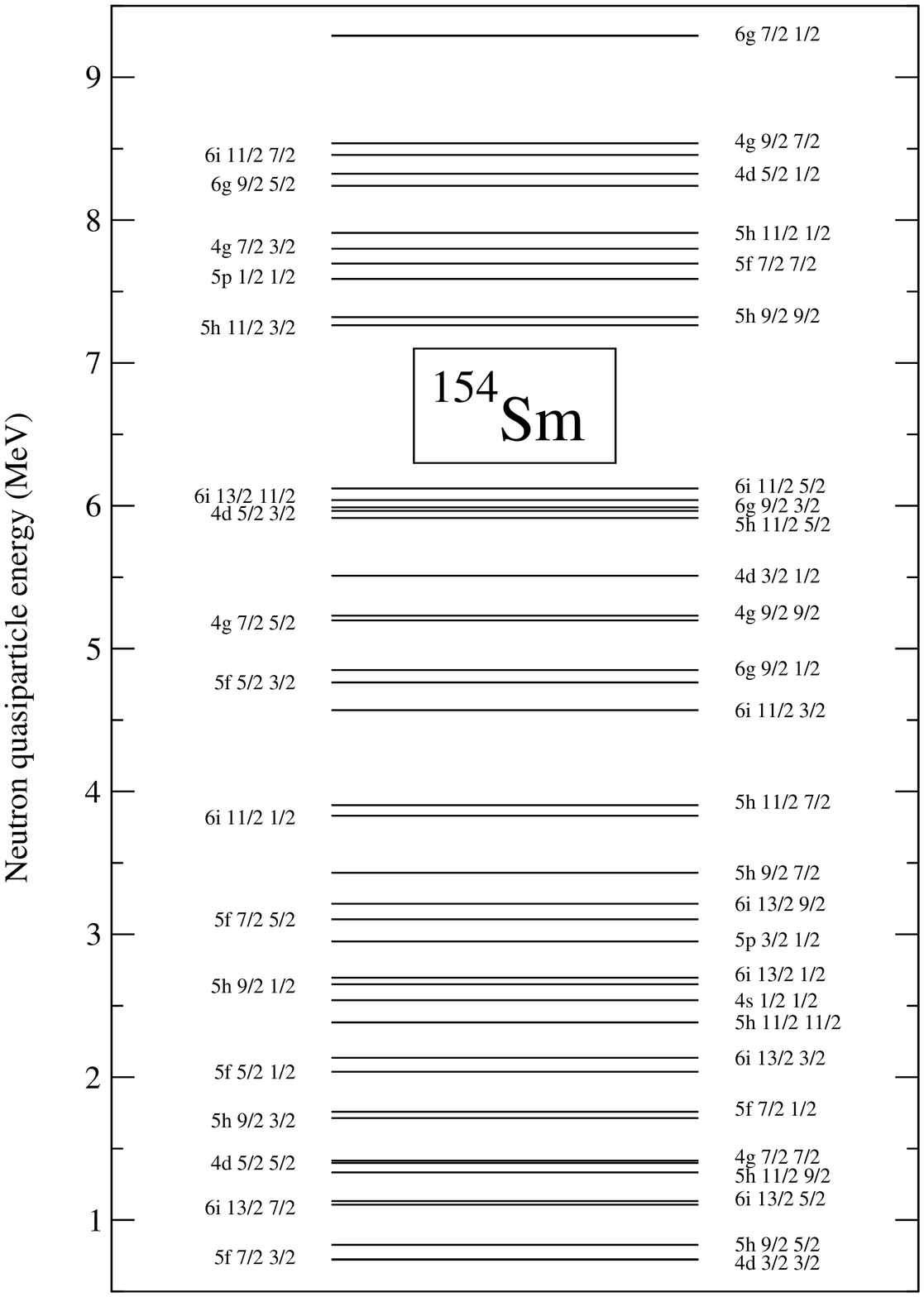,width=10cm,bbllx=5cm,%
bblly=10cm,bburx=18cm,bbury=26cm,angle=0}}
\vskip8cm
\caption{Neutron quasiparticle energies given in units of MeV.}
\label{Fig. 4}
\end{figure}
\clearpage

\begin{figure}[h]
\centerline{\psfig{figure=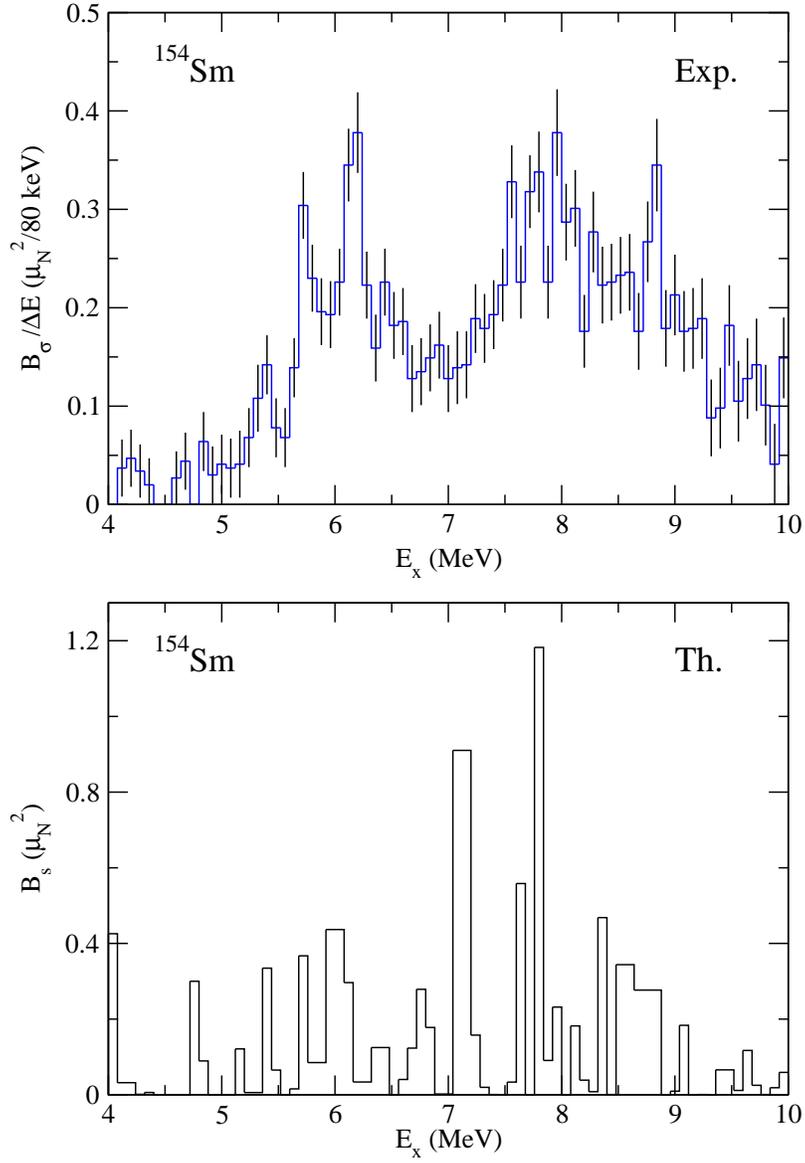,width=8cm,bbllx=5cm,%
bblly=10cm,bburx=18cm,bbury=26cm,angle=0} }
\vskip8cm
\caption{The theoretical histogram of the spin strength (lower panel) is
compared with the experimental strength distribution. 
The experimental plot is taken from Ref.[9]}
\label{Fig. 5}
\end{figure}
\clearpage

\begin{figure}[h]
\centerline{\psfig{figure=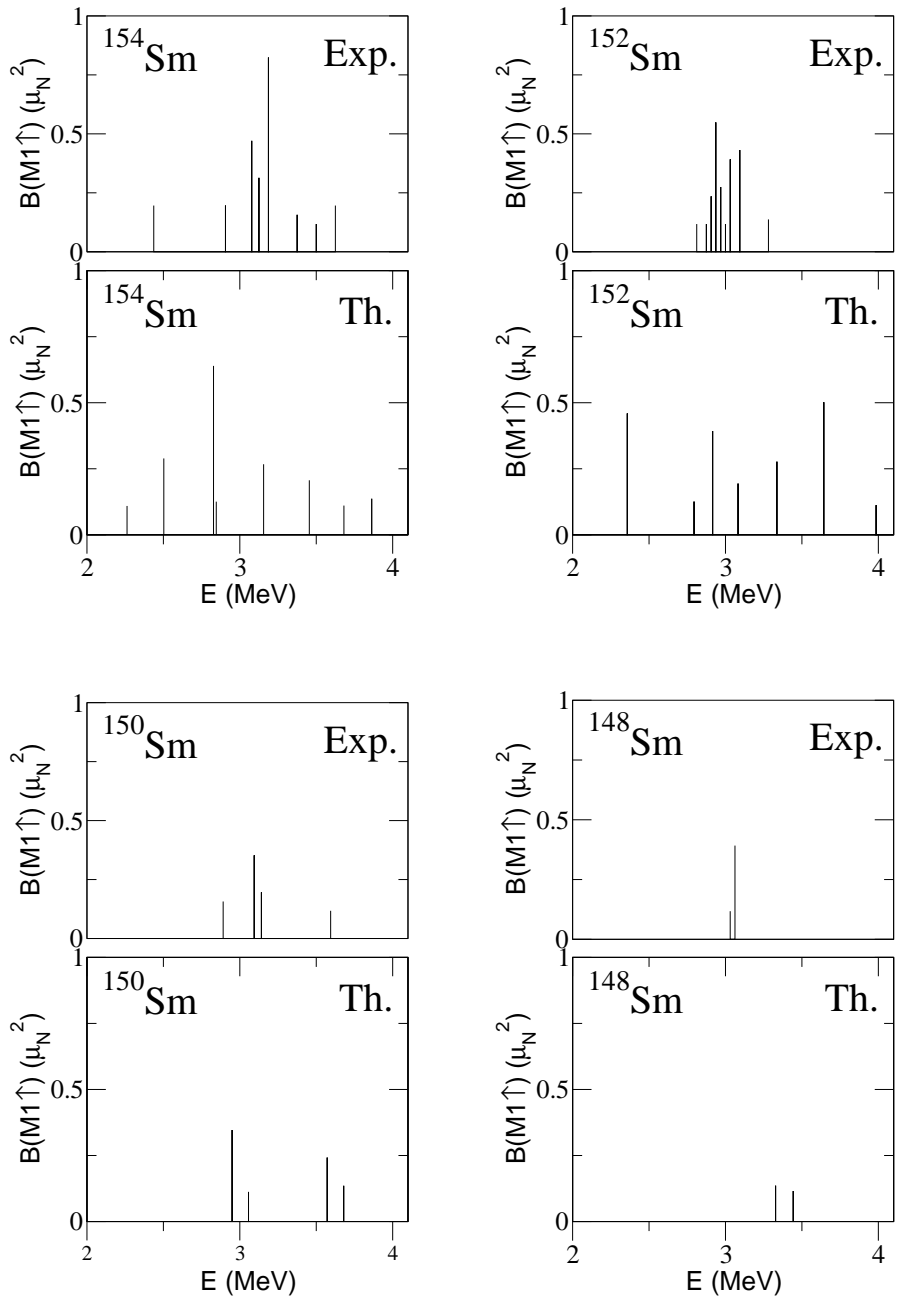,width=8cm,bbllx=5cm,%
bblly=10cm,bburx=18cm,bbury=26cm,angle=0} }
\vskip8cm
\caption{The predicted orbital strength (lower part) is compared with the experimental data (upper part). Experimental data are from Refs[2-9]}
\label{Fig. 6}
\end{figure}
\clearpage

\begin{figure}[h]
\centerline{\psfig{figure=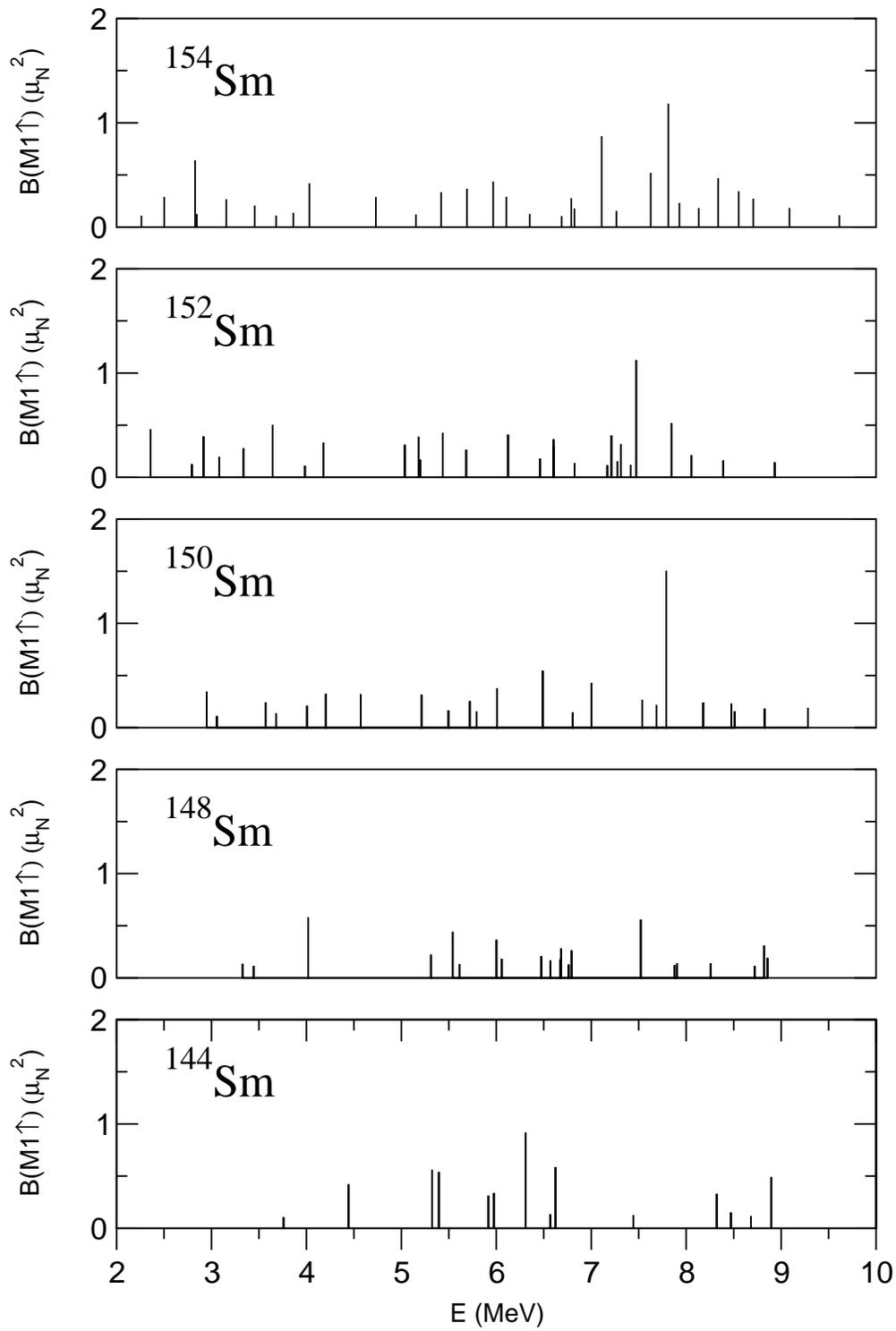,width=10cm,bbllx=5cm,%
bblly=10cm,bburx=18cm,bbury=26cm,angle=0} }
\vskip8cm
\caption{Theoretical M1 strengths, in the range of 2 to 10 MeV, 
are plotted as function of energy for some even-even Sm isotopes. }
\label{Fig. 7}
\end{figure}
\clearpage

\begin{figure}[h]
\centerline{\psfig{figure=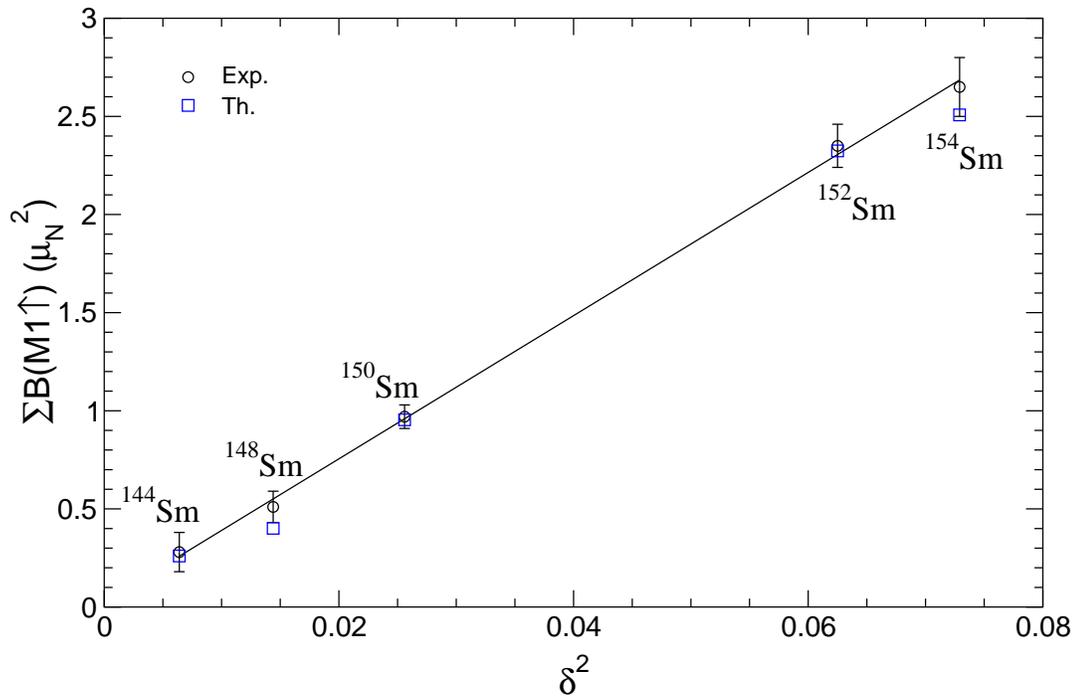,width=8cm,bbllx=3cm,%
bblly=10cm,bburx=18cm,bbury=26cm,angle=-90} }
\vskip9cm
\caption{The orbital total strength in the range 0-4 MeV is presented as function of $\delta^2$, for the even-even Sm isotopes. Both the theoretical and experimental results (Refs. [2-8]) are shown.}
\label{Fig. 8}
\end{figure}
\clearpage

\begin{figure}[h]
\centerline{\psfig{figure=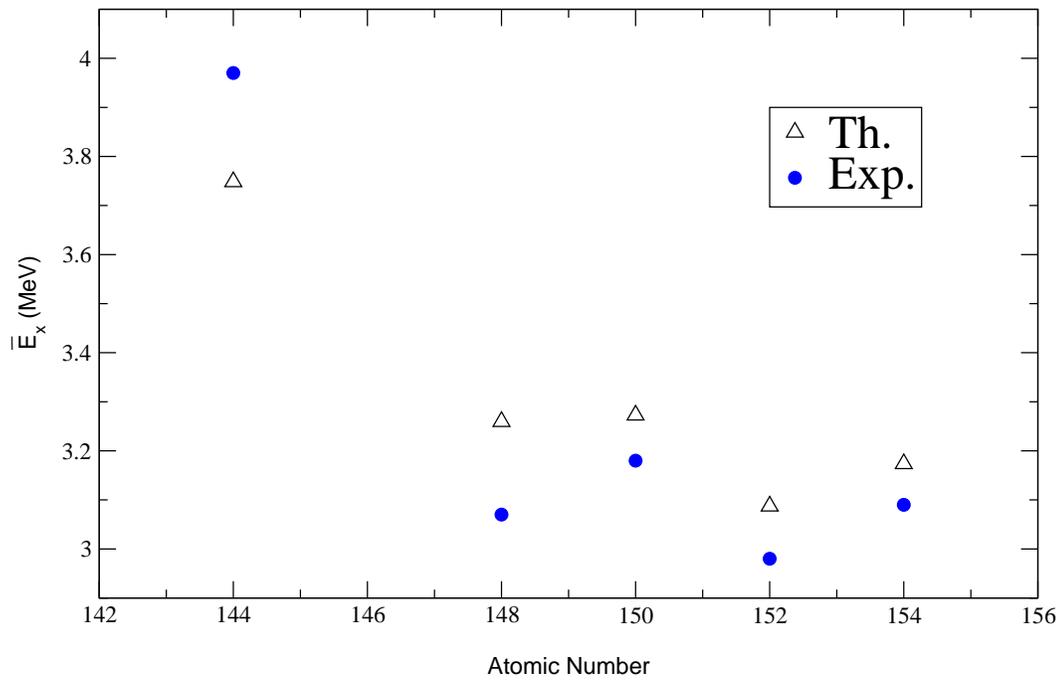,width=8cm,bbllx=3cm,%
bblly=10cm,bburx=18cm,bbury=26cm,angle=-90} }
\vskip9cm
\caption{Theoretical (triangles) and experimental (black circles) average energies calculated by means of Eq. (4.3)are given for the 5 isotopes of Sm, studied
in the present paper.}
\label{Fig. 9}
\end{figure}
\clearpage

\begin{figure}[h]
\centerline{\psfig{figure=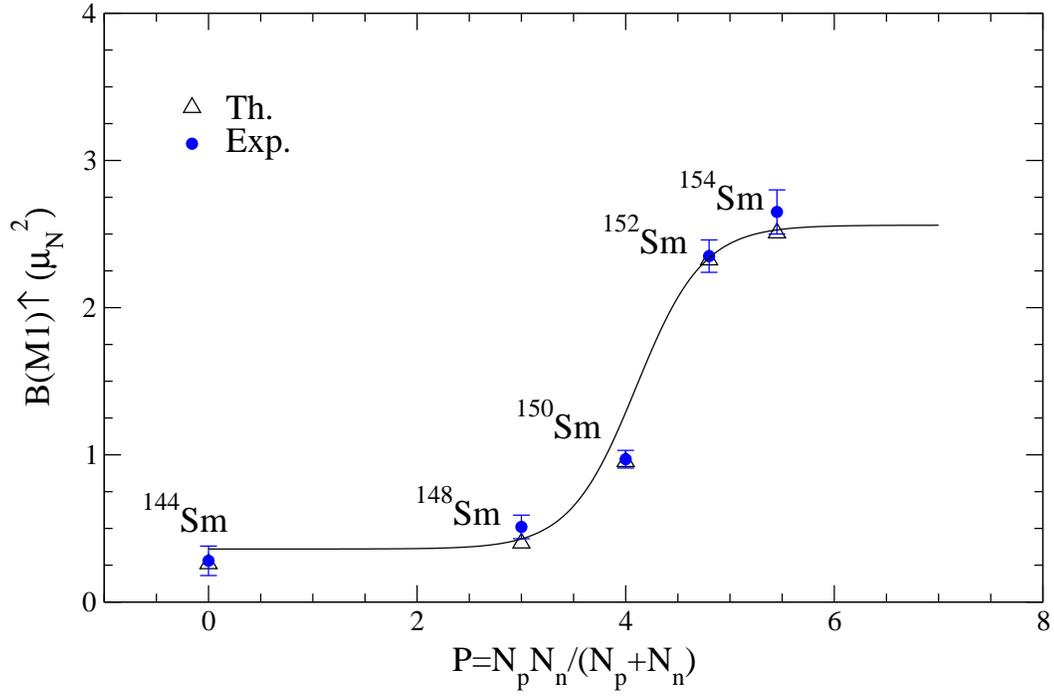,width=8cm,bbllx=3cm,%
bblly=10cm,bburx=18cm,bbury=26cm,angle=-90} }
\vskip9cm
\caption{The orbital M1 strength in the interval 0-4 MeV, as function of the ratio P defined by Eq. (4.4).}
\label{Fig. 10}
\end{figure}
\clearpage

\end{document}